# What does a violation of the Bell's inequality prove?

Sergey A. Rashkovskiy

*Institute for Problems in Mechanics of the Russian Academy of Sciences, Vernadskogo Ave., 101/1, Moscow, 119526, Russia*

*Tomsk State University, 36 Lenina Avenue, Tomsk, 634050, Russia*

*E-mail: rash@ipmnet.ru, Tel. +7 906 0318854*

## ABSTRACT

In this paper I show that the Einstein-Podolsky-Rosen-Bohm Gedankenexperiment and so-called entanglement of photons have a simple explanation within the framework of classical electrodynamics if we take into account the discrete (atomic) structure of the detectors and a specific nature of the light–atom interaction. In this case we do not find such a paradox as "spooky action at a distance". I show that CHSH criterion in EPRB Gedankenexperiment with classical light waves can exceed not only a maximum value $S_{HV} = 2$ which is predicted by the local hidden-variable theories but also the maximum value $S_{QM} = 2\sqrt{2}$ predicted by quantum mechanics and in this case there is no desire to construct a local hidden-variable theory.

## I. INTRODUCTION

One of the most mysterious and intriguing predictions of quantum mechanics is the entanglement phenomenon, which manifests in a strong correlation of the behavior of quantum objects, even when they are separated by a large distance. According to quantum mechanics, the state of each such an object cannot be described independently – instead, a quantum state must be described for the system as a whole. The entangled state cannot be factorized into a product of two states associated to each object. According to this, it is considered that we cannot ascribe any well-defined state to each object.

The entanglement phenomenon is considered to be a basis for new hypothetical solutions, primarily in the field of information technologies.

This phenomenon was considered for the first time by A. Einstein, B. Podolsky, and N. Rosen [1] and was developed further by D. Bohm [2] who described what came to be known as the EPRB Gedankenexperiment and EPRB paradox.

The first quantitative criterion which describes such a paradox was proposed by J. Bell (Bell's inequality) [3]. The Bell's inequality, derived on the basis of the local hidden-variable theories, contradicts in some cases the predictions of quantum mechanics. It is considered that an experiment in which the violation of the Bell's inequality occurs cannot be explained based on the local realism view. Bell's inequality gave the tool for experimental verification of the counterintuitive predictions of quantum mechanics. Later, J. F. Clauser, M.A. Horne, A. Shimony and R.A. Holt (CHSH) proposed a new criterion and an experiment to test the local hidden-variable theories [4].

In these experiments, the two photons $v_1$ and $v_2$, emitted in the entangled state, are analyzed by linear polarizers in orientations **a** and **b** [5-7] (Fig. 1). Each polarizer is followed by two detectors, giving results + or −, corresponding to a linear polarization found parallel or perpendicular to **a** and **b**.

Measuring the clicks of the detectors one can calculate the probabilities of events, both singles and their coincidences.

Quantum mechanics predicts for singles probabilities

$$P_+(\mathbf{a}) = P_-(\mathbf{a}) = P_+(\mathbf{b}) = P_-(\mathbf{b}) = 1/2 \qquad (1)$$

where $P_\pm(\mathbf{a})$ and $P_\pm(\mathbf{b})$ are the probabilities of getting the results ± for the photons $v_1$ and $v_2$ respectively.

These results are in agreement with that each individual polarization measurement gives a random result and with the point of view that the photon is indivisible and we cannot observe simultaneously the clicks of the detectors $\mathbf{a}_+$ and $\mathbf{a}_-$ for polarizer **a** and correspondingly the clicks of the detectors $\mathbf{b}_+$ and $\mathbf{b}_-$ for polarizer **b**. Accordingly to [3,4] the entanglement of the photons manifests in the probabilities $P_{\pm\pm}(\mathbf{a},\mathbf{b})$ of joint detections of $v_1$ and $v_2$ in the channels + or − of polarizers **a** and **b**. For entangled particles, quantum mechanics predicts:

$$\begin{aligned} P_{++}(\mathbf{a},\mathbf{b}) &= P_{--}(\mathbf{a},\mathbf{b}) = \frac{1}{2}\cos^2\alpha \\ P_{+-}(\mathbf{a},\mathbf{b}) &= P_{-+}(\mathbf{a},\mathbf{b}) = \frac{1}{2}\sin^2\alpha \end{aligned} \qquad (2)$$

where $\alpha$ is the angle between orientations of the polarizers **a** and **b.**

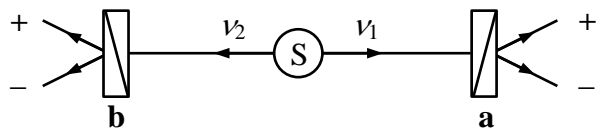

FIG. 1. Einstein-Podolsky-Rosen-Bohm Gedankenexperiment with photons [4-7].

In order to describe quantitatively the correlations between random events, one can introduce the correlation coefficient [4]

$$E(\mathbf{a},\mathbf{b}) = P_{++}(\mathbf{a},\mathbf{b}) - P_{+-}(\mathbf{a},\mathbf{b}) - P_{-+}(\mathbf{a},\mathbf{b}) + P_{--}(\mathbf{a},\mathbf{b}) \tag{3}$$

Using (2), quantum mechanics predicts

$$E_{QM}(\mathbf{a},\mathbf{b}) = \cos 2\alpha \tag{4}$$

Carrying out the experiments for four different orientations $\mathbf{a}, \mathbf{a}'$ and $\mathbf{b}, \mathbf{b}'$ of the polarizers $\mathbf{a}$ and $\mathbf{b}$, one can calculate the parameter [4]

$$S = E(\mathbf{a},\mathbf{b}) - E(\mathbf{a},\mathbf{b}') + E(\mathbf{a}',\mathbf{b}) + E(\mathbf{a}',\mathbf{b}') \tag{5}$$

The local hidden-variable theories predict [4]

$$-2 \leq S_{HV} \leq 2 \tag{6}$$

It is well known that the greatest conflict between quantum mechanical predictions and CHSH inequalities (6) that follows from the local hidden-variable theories [4] is expected for the set orientations $(\mathbf{a},\mathbf{b}) = (\mathbf{a}',\mathbf{b}) = (\mathbf{a}',\mathbf{b}') = 22.5^{\mathrm{o}}$ and $(\mathbf{a},\mathbf{b}') = 67.5^{\mathrm{o}}$. In this case, quantum mechanics predicts

$$S_{QM} = 2\sqrt{2} \tag{7}$$

CHSH inequality (6) was testable in numerous relevant experiments, starting with the pioneering works [5-7], all of which have shown agreement with quantum mechanics rather than the principle of local realism. Violation of Bell's inequalities (6) was fixed for a wide range of the distances and timings of the measurements [8-16].

These results have shown, in particular, that the EPRB experiments with "entangled photons" cannot be described within the local hidden-variable theories and in general that it is impossible to construct the local hidden-variable theories which are capable to describe the quantum mechanical regularities. From this point of view, the entanglement is considered as a direct evidence of the existence of photons.

The paradox of the results [5-16] is that these experiments can be physically explained based on the photonic (corpuscular) representations only if one assumes that the interaction between the two entangled particles and between the particles and measuring devices are propagated at a velocity substantially exceeding the speed of light, which contradicts the relativity theory.

Einstein characterized it as "spooky action at a distance" and argued that the accepted formulation of quantum mechanics must therefore be incomplete.

We note that these conclusions are based on the photonic representations, i.e., on the fact that each click of the detector is associated with a hit of a particle - a photon.

Let us recall that the coincidence experiments were started with the pioneering Hanbury Brown and Twiss experiments [17,18] the results of which initially also have caused surprise. Later, the simple explanation of the Hanbury Brown and Twiss effect was found within the framework of semiclassical theory without quantization of radiation [19,20]. The idea of this explanation is as follows.

Solution of the Schrödinger equation allows calculating the probability of excitation of an atom of the detector by classical electromagnetic (light) wave for time $\Delta t$ (Fermi's golden rule)

$$w\Delta t = bI\Delta t \qquad (8)$$

where $w$ is the probability of excitation of atoms per unit time; $I \sim \mathbf{E}^2$ is the intensity of the classical light wave at the location of the atom; and $b$ is a constant which does not depend on the intensity of incident light. In this case, each click of the detector is considered to be the result of the excitation of one of the atoms under the action of light. Assuming that the components of the electric field vector of the light wave $\mathbf{E}$ are random variables and have a Gaussian distribution, one can calculate all regularities of the Hanbury Brown and Twiss effect [19,20]. The Hanbury Brown and Twiss correlation appears as a result of correlation of intensities of light waves arriving at the two detectors due to splitting the incident light wave.

As shown in [21-23], the experiments with so-called "single photons" (but actually with the weak classical light waves), namely the double-slit experiments and the Wiener experiments with standing light waves, can completely be described using the expression (8) within the framework of semiclassical theory without quantization of radiation.

Based on the results [19-23], one can conclude that the so-called "wave–particle duality" of light has a simple explanation within the framework of classical electrodynamics if we take into account the discrete (atomic) structure of detector, while the concept of a "photon" becomes superfluous.

In this paper I show that the Einstein-Podolsky-Rosen-Bohm Gedankenexperiment and so-called entanglement of "photons" have also a simple explanation within the framework of classical electrodynamics if we take into account the expression (8) and atomic structure of the detectors.

## II. EPRB GEDANKENEXPERIMENT WITH CLASSICAL LIGHT WAVES

Let us consider the EPRB Gedankenexperiment with classical light waves (Fig. 2). We assume that a source S emits two identical classical electromagnetic (light) waves in two opposite directions; i.e.

$$\mathbf{E}_1 = \mathbf{E}_2 = \mathbf{E} \qquad (9)$$

for light waves $\nu_1$ and $\nu_2$ (Fig. 2).

The emitted waves $v_1$ and $v_2$ arrive at two spatially separated two-channel polarizers **a** and **b**, each of which splits the incident light beam into two mutually orthogonal linearly polarized beams that arrive at corresponding detectors. For each polarizer, we introduce its own coordinate system $(x, y)$, where the $x$ axis is parallel to the axis of the polarizer, while the axis $y$ is perpendicular to it. Further, the beam with polarization parallel to the axis of the polarizer is denoted by the index "+", while the beam with polarization perpendicular to the axis of the polarizer is denoted by the index "−". Corresponding detectors will be denoted as $a_\pm$ and $b_\pm$.

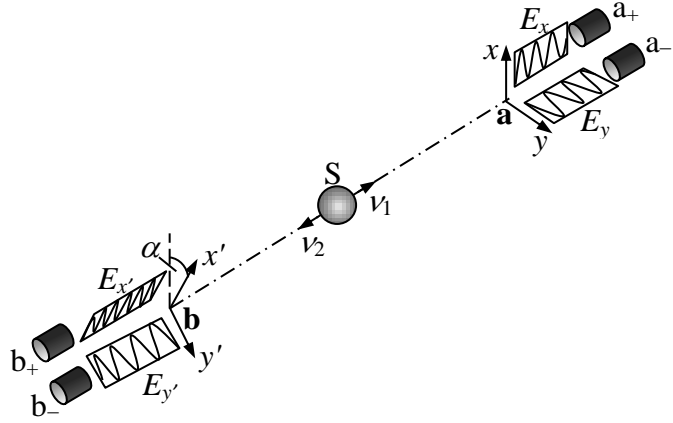

FIG. 2. Einstein-Podolsky-Rosen-Bohm Gedankenexperiment with classical light waves.

Each polarizer can rotate around the axis of the incident beam. Due to the isotropy of the system, only a relative angle of rotation of the polarizers $\alpha$ has the meaning. Therefore, we choose a coordinate system associated with polarizer **a** as the reference system. Furthermore, we assume that polarizers are ideal, i.e. we neglect the energy losses of light wave at passing the optical system. Then only the component $E_x$ of the incident light wave $v_1$ will arrive at the detector $a_+$ while only the component $E_y$ of this wave will arrive at the detector $a_-$. The intensities of light waves arriving at the detectors $a_\pm$ are as follows

$$I_+(\mathbf{a}) = E_x^2 \,;\; I_-(\mathbf{a}) = E_y^2 \qquad (10)$$

Let polarizer **b** is turned with respect to the polarizer **a** at an angle $\alpha$ (Fig. 2). We denote the own coordinate system of the polarizer **b** as $(x', y')$, whose axes are parallel to the corresponding main axis of the polarizer **b**. Then only the component $E_{x'}$ of the incident light wave $v_2$ will arrive at the detector $b_+$ while only the component $E_{x'}$ of this wave will arrive at the detector $b_-$.

Taking into account (9), one can write

$$E_{x'} = E_x \cos\alpha + E_y \sin\alpha, \quad E_{y'} = -E_x \sin\alpha + E_y \cos\alpha \tag{11}$$

The intensities of light waves arriving at the detectors $b_\pm$ are, respectively,

$$I_+(\mathbf{b}) = E_{x'}^2; \quad I_-(\mathbf{b}) = E_{y'}^2 \tag{12}$$

Under the action of the incident light wave, the excitation of the atoms of a detector can occur. We assume that the excitation of an atom of the detector inevitably causes an electron avalanche in the detector, which manifests in the form of a single event (click of detector), which is fixed by the registrar.

It is believed that after the triggering, the detector (an atom) returns again into the initial (ground) state and ready for the next act of excitation.

The rate of atomic excitation $w$ is described by the expression (8) and does not depend on the concentration of atoms. If the radiation intensity does not change within the time of exposure (within a time window), then the law of excitation of atoms will be similar to that of radioactive decay. In particular, the probability of excitation of an atom during a time $\Delta t$ is [21-23]

$$P(t) = 1 - \exp(-w\Delta t) \tag{13}$$

Taking into account (8), one obtains

$$P(t) = 1 - \exp(-bI\Delta t) \tag{14}$$

where $I$ is the intensity of light wave arriving at a corresponding detector.

Assuming that a source of radiation is Gaussian, and that the components of the electric field vector $\mathbf{E} = (E_x, E_y)$ of the light wave are statistically independent, one obtains the probability density for the components of the electric field vector

$$p(E_x, E_y) = \frac{1}{2\pi I_0} \exp\left(-\frac{E_x^2 + E_y^2}{2I_0}\right) \tag{15}$$

where

$$I_0 = \langle E_x^2 \rangle = \langle E_y^2 \rangle \tag{16}$$

$\langle ... \rangle$ denotes averaging.

Obviously, considering (9),

$$\langle E_{x'}^2 \rangle = \langle E_{y'}^2 \rangle = I_0 \tag{17}$$

Equations (14)-(17) allow calculating the EPRB Gedankenexperiment in detail. Indeed, calculating the single events of triggering the detectors $a_\pm$ and $b_\pm$ using the expressions (14) and (15), one can determine their statistics both for single events and for their coincidences and compare it with the predictions of quantum mechanics (1), (2).

## III. MONTE CARLO SIMULATION OF EPRB GEDANKENEXPERIMENT

Let us consider the discrete time intervals (time windows) $i = 1,2,...,N$ which have a duration $\Delta t$, during which, the discrete events occurring at different detectors are recorded. The events on different detectors will be considered as simultaneous if they occurred within the same time window $i$. At the same time, the events occurring at different detectors are statistically independent, and are described by the probabilities (14), in which we use the intensities (10) and (12) of the light waves arriving at corresponding detector within a given time window. The intensity of the light waves $v_1$ and $v_2$ emitted by a source for different time windows are considered to be random and are described by the probability density function (15).

Let us introduce the nondimensional exposure time (nondimensional duration of time window)

$$\tau = bI_0\Delta t \qquad (18)$$

In this case, the probability of excitation of the atom during a time window is

$$P(t) = 1 - \exp\left(-(I/I_0)\tau\right) \qquad (19)$$

Further, we take the value $I_0$ as a characteristic intensity of light. In this case we use the parameter $I_0^{1/2}$ as a scale for the field $\mathbf{E}$. Taking into account the expressions (15) and (19), in further calculations we will take $I_0 = 1$, while the value $I_0$ itself will be included into nondimensional duration $\tau$ of time window.

Then the expressions (15) and (19) can be written in nondimensional form:

$$P(\tau) = 1 - \exp(-I\tau) \qquad (20)$$

$$p(E_x, E_y) = \frac{1}{2\pi}\exp\left(-\frac{E_x^2 + E_y^2}{2}\right) \qquad (21)$$

where $I = \mathbf{E}^2$.

Thus in the problem under consideration, there is a single nondimensional parameter $\tau$ varying which we can change the "experimental conditions".

The calculation proceeds quite trivial using the Monte Carlo method: at each time window $i$, the components of the electric field vector $\mathbf{E}$ of the light wave are generated using the probability density (21). Using the components $E_x$ and $E_y$, the intensities of the light waves (10) and (12) arriving at the corresponding detector are calculated taking into account the expression (11). Using these intensities, the probabilities of excitation of each detector are calculated using the expression (20). At the same time the random numbers $R \in [0,1]$ are generated for each detector using a random number generator. If the condition $R \leq P$ is satisfied for some detector, it is considered to be excited and this event is recorded in corresponding time window. Thus, we record all events of triggering the detectors at different time windows. Note that in these calculations, the assumption was made that no more than one discrete event can occur at one detector within one time window. In real experiments [5-16], the duration of the time window was significantly longer then relaxation time of the detector. Therefore, generally speaking, there is a finite probability that the same detector will trigger several times during the same time window. Accounting for no more than one triggering the same detector within the same time window, in fact, means the rejection of such time windows in a real experiment.

After all time windows $i = 1,2,...,N$ were calculated, the statistical analysis both the single events, and the coupled events (coincidences) for different pairs of detectors $a_\pm$ and $b_\pm$ is performed. This allows determining any statistical characteristics of such an experiment.

For us, it is of interest to analyze the results of Monte Carlo simulations based on photonic (corpuscular) representations. For this purpose, we will assume that each triggering the detector is the result of hit on it a particle - a photon. At the same time, we will remember that in reality, the results of Monte Carlo simulation were obtained within the framework of semiclassical theory, in which light is considered as a classical electromagnetic wave, while the photonic model is just a fiction, the goal of which is a mechanistic (naive) "explanation" of discrete events of triggering the detectors.

As soon as we begin to analyze the results of experiments from the standpoint of the photonic representations, we immediately have to introduce a number of limitations related to our ideas about photons as indivisible particles.

First of all, if both detectors behind the same polarizer simultaneously triggered during one time window, for indivisible photon, such an event can be explained either by a background or by interferences in circuit, or by an error in the detector operation. In any case, this result leads to violation of the conditions (1), and therefore the time windows in which such events occurred, should be rejected.

Further, assuming that the source S emits a pair of entangled photons, we can expect that within one time window, the simultaneous triggering the detectors behind the both polarizers will be recorded. In other words, if a photon was detected behind the polarizer **a**, then the second photon must also be detected behind the polarizer **b**. If a second event did not happen, then the result can be explained either by an insufficient sensitivity of a detector, or by a malfunction of a detector, or by the fact that a one photon of the entangled pair was "lost" on the way to a polarizer, which is also perceived as an detection error and such time windows should not be taken into account when calculating the "photon coincidences."

Thus when counting the number of coincidences we have to reject not only the time windows in which both detectors triggered behind any polarizer, but also those time windows at which one detector triggered behind one of polarizers while there are no detectors triggered behind other polarizer, because only such events are consistent with the "photonic representations" in this experiment.

Thus, by statistical analysis of the results of semiclassical Monte Carlo simulations of EPRB Gedankenexperiment with classical light waves, we calculate the number of the time windows in which the pairs of corresponding events have occurred: $N_{++}(\mathbf{a},\mathbf{b})$, $N_{--}(\mathbf{a},\mathbf{b})$, $N_{+-}(\mathbf{a},\mathbf{b})$, $N_{-+}(\mathbf{a},\mathbf{b})$. For example, $N_{++}(\mathbf{a},\mathbf{b})$ is the number of time windows in which the events were recorded simultaneously on the detectors $a_+$ and $b_+$, while the events on other detectors were not observed, etc. Then

$$N_0 = N_{++}(\mathbf{a},\mathbf{b}) + N_{--}(\mathbf{a},\mathbf{b}) + N_{+-}(\mathbf{a},\mathbf{b}) + N_{-+}(\mathbf{a},\mathbf{b}) \qquad (22)$$

is the number of the time windows at which the events were recorded behind both polarizers but only one detector has triggered behind each polarizer.

Then the probabilities of the corresponding pair events are determine by the expression

$$P_{\pm\pm}(\mathbf{a},\mathbf{b}) = N_{\pm\pm}(\mathbf{a},\mathbf{b})/N_0 \qquad (23)$$

The probabilities (23) determined exactly in this way correspond to those calculated in quantum mechanics.

Using the probabilities (23) for each relative orientation $\alpha$ of the polarizers one can calculate the correlation coefficient (3).

The results of Monte Carlo simulations of the EPRB Gedankenexperiment with classical light waves for different values of nondimensional width $\tau$ of time window processed statistically based on the photonic representations are shown in Figs. 3-5. Ibid, the dependences (2) and (4) predicted by quantum mechanics are shown.

We can see that at large $\tau \gg 1$, the results of semiclassical Monte Carlo simulations practically coincide with the predictions of quantum mechanics (2) and (4) for entangled photons. In particular, the probabilities $P_{++}(\alpha = 0)$ and $P_{--}(\alpha = 0)$ differ slightly from 0.5, which in a real EPRB experiment, could be attributed due to a non-ideality of the optical equipment, as it was done in [5-7].

At the same time, we see that the probabilities $P_{\pm\pm}(\mathbf{a},\mathbf{b})$ and the correlation coefficient $E(\mathbf{a},\mathbf{b})$ are increasingly deviated from the predictions of quantum mechanics (2) and (4) with the decrease of nondimensional width $\tau$ of time window.

Thus, in the EPRB Gedankenexperiment with classical light waves, we observe exactly the effect which is called entanglement of photon. We see that entanglement is observed at $\tau \gg 1$ only after statistical processing the "experimental data" based on the photonic representations and is related to nonlinear dependence (20) at $\tau \gg 1$.

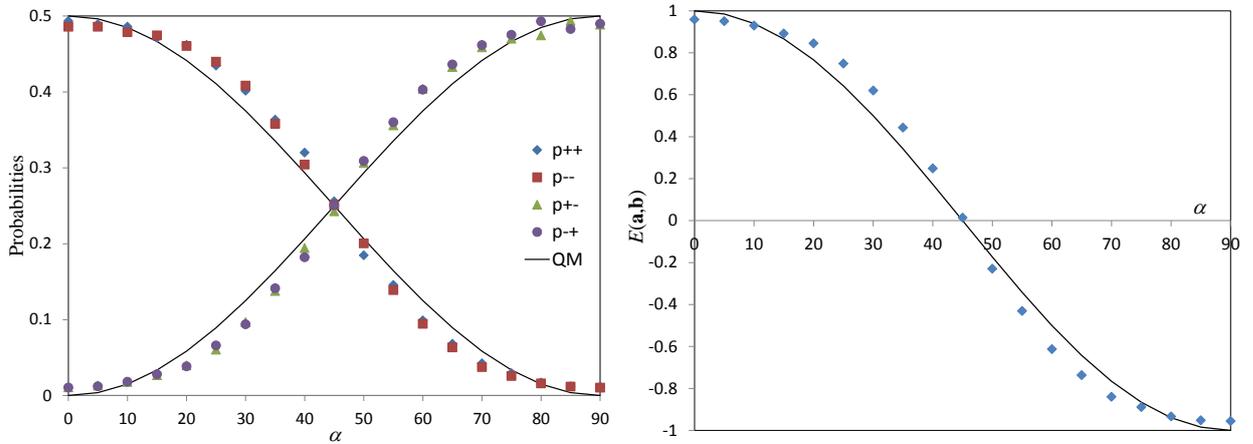

FIG. 3. (Colour online) Dependence of the probabilities of the pairs of events (left) and correlation coefficient $E(\mathbf{a},\mathbf{b})$ (right) on the angle between polarizers for $\tau = 20$. Markers are the results of semiclassical Monte Carlo simulations; lines are the predictions of quantum mechanics (2) and (4).

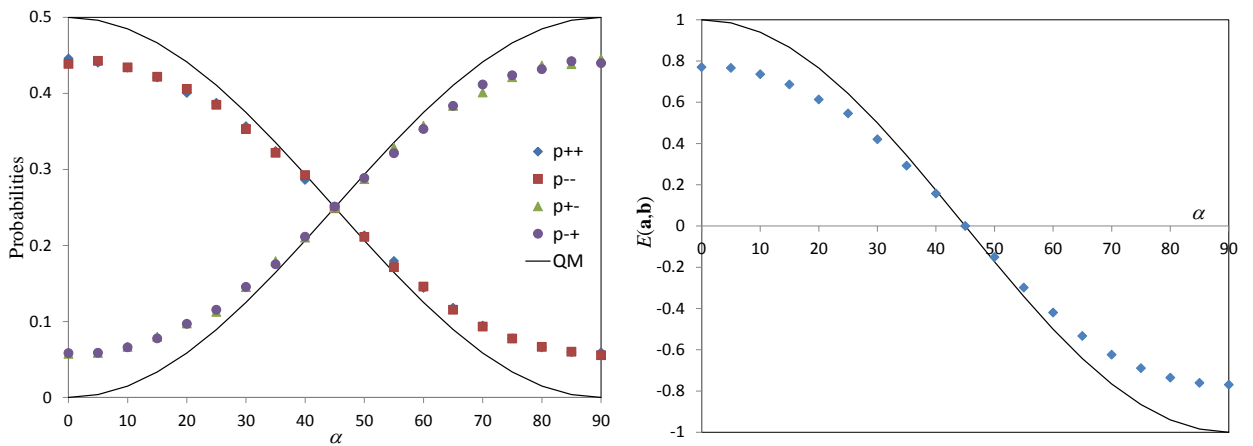

FIG. 4. (Colour online) Dependence of the probabilities of the pairs of events (left) and correlation coefficient $E(\mathbf{a},\mathbf{b})$ (right) on the angle between polarizers for $\tau = 1$. Markers are the results of semiclassical Monte Carlo simulations; lines are the predictions of quantum mechanics (2) and (4).

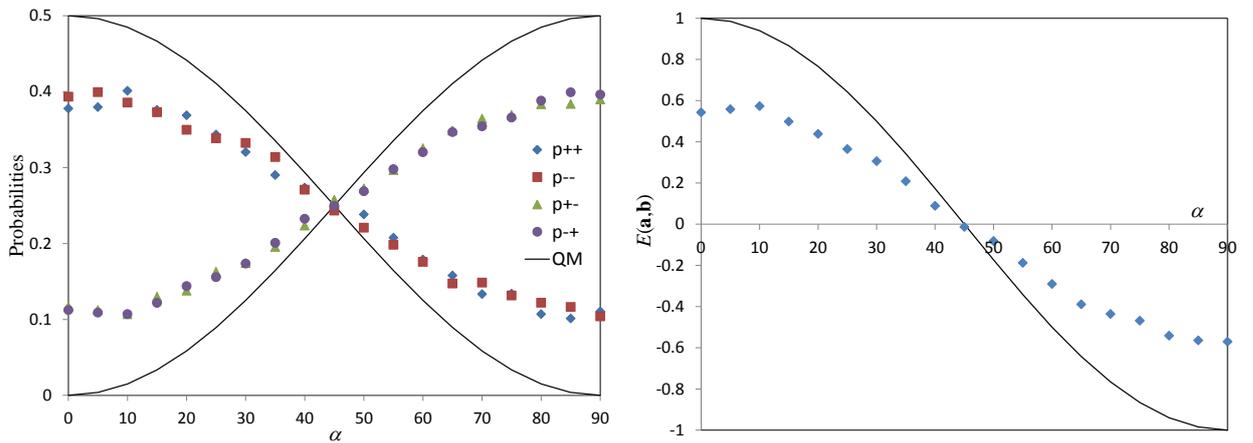

FIG. 5. (Colour online) Dependence of the probabilities of the pairs of events (left) and correlation coefficient $E(\mathbf{a},\mathbf{b})$ (right) on the angle between polarizers for $\tau =0.1$. Markers are the results of semiclassical Monte Carlo simulations; lines are the predictions of quantum mechanics (2) and (4).

## IV. ANALYTICAL DESCRIPTION OF EPRB GEDANKENEXPERIMENT

Let us obtain the expressions for probabilities $P_{\pm\pm}(\mathbf{a},\mathbf{b})$.

First of all, we note that in the experiment under consideration, the splitting the light beam at the one polarizer is equivalent to the Hanbury Brown and Twiss experiment, with the only difference being that here the polarizer selects two mutually perpendicular, and thus statistically independent components of the vector $\mathbf{E}$: $\langle E_x^2 E_y^2 \rangle = \langle E_x^2 \rangle \langle E_y^2 \rangle = I_0^2$; at the same time, in the Hanbury Brown and Twiss experiment, each light beam behind a splitter is a mixture of both polarizations, and therefore both the beams behind a splitter are correlated: for example, for a Gaussian beam $\langle I_1 I_2 \rangle = 2 \langle I_1 \rangle \langle I_2 \rangle$.

We first calculate $N_{++}(\mathbf{a},\mathbf{b})$ in EPRB Gedankenexperiment with classical light waves.

Due to independence of the events on each detector, the probability that, at fixed $E_x$ and $E_y$, the clicks of both detectors $\mathbf{a}_+$ and $\mathbf{b}_+$ will occur simultaneously but at the same time the clicks of detectors $\mathbf{a}_-$ and $\mathbf{b}_-$ will not occur is equal to $P_+(\mathbf{a})P_+(\mathbf{b})[1-P_-(\mathbf{a})][1-P_-(\mathbf{b})]$, where

$$P_\pm(\mathbf{a}) = 1 - \exp(-I_\pm(\mathbf{a})\tau);\ P_\pm(\mathbf{b}) = 1 - \exp(-I_\pm(\mathbf{b})\tau) \qquad (24)$$

are the probabilities of triggering the corresponding detectors behind the polarizers $\mathbf{a}$ and $\mathbf{b}$.

Then, averaging over the all possible realizations of the parameters $E_x$ and $E_y$, one obtains

$$N_{++}(\mathbf{a},\mathbf{b}) = N \langle P_+(\mathbf{a}) P_+(\mathbf{b})[1-P_-(\mathbf{a})][1-P_-(\mathbf{b})] \rangle \qquad (25)$$

where $N$ is the total number of the time windows; the averaging is carried out using the probability density (21):

$$N_{++}(\mathbf{a},\mathbf{b}) = N \int_{-\infty}^{\infty} \int_{-\infty}^{\infty} P_+(\mathbf{a}) P_+(\mathbf{b}) [1 - P_-(\mathbf{a})][1 - P_-(\mathbf{b})] p(E_x, E_y) dE_x dE_y \quad (26)$$

Taking (21) and (24) into account, after the simple calculations one obtains

$$N_{++}(\mathbf{a},\mathbf{b})/N = \frac{1}{\sqrt{1 + 4\tau + 4\tau^2 \sin^2\alpha}} - \frac{2}{\sqrt{1 + 6\tau + 8\tau^2}} + \frac{1}{(1+4\tau)} \quad (27)$$

Similarly

$$N_{+-}(\mathbf{a},\mathbf{b}) = N \int_{-\infty}^{\infty} \int_{-\infty}^{\infty} P_+(\mathbf{a}) P_-(\mathbf{b}) [1 - P_-(\mathbf{a})][1 - P_+(\mathbf{b})] p(E_x, E_y) dE_x dE_y \quad (28)$$

and after the simple calculations one obtains

$$N_{+-}(\mathbf{a},\mathbf{b})/N = \frac{1}{\sqrt{1 + 4\tau + 4\tau^2 \cos^2\alpha}} - \frac{2}{\sqrt{1 + 6\tau + 8\tau^2}} + \frac{1}{(1+4\tau)} \quad (29)$$

It is easy to show that

$$N_{++}(\mathbf{a},\mathbf{b}) = N_{--}(\mathbf{a},\mathbf{b}), \quad N_{-+}(\mathbf{a},\mathbf{b}) = N_{+-}(\mathbf{a},\mathbf{b}) \quad (30)$$

Then for the conditional probabilities (23) one obtains the expression

$$P_{\pm\pm}(\mathbf{a},\mathbf{b}) = \frac{N_{\pm\pm}(\mathbf{a},\mathbf{b})}{2(N_{++}(\mathbf{a},\mathbf{b}) + N_{+-}(\mathbf{a},\mathbf{b}))} \quad (31)$$

Obviously, the normalization of probabilities takes place:

$$P_{++}(\mathbf{a},\mathbf{b}) + P_{--}(\mathbf{a},\mathbf{b}) + P_{+-}(\mathbf{a},\mathbf{b}) + P_{-+}(\mathbf{a},\mathbf{b}) = 1 \quad (32)$$

Taking (30) into account one obtains

$$P_{++}(\mathbf{a},\mathbf{b}) = P_{--}(\mathbf{a},\mathbf{b}), \quad P_{-+}(\mathbf{a},\mathbf{b}) = P_{+-}(\mathbf{a},\mathbf{b}) \quad (33)$$

Using the probabilities (31) and expression (3), it is easy to calculate the correlation coefficient $E(\mathbf{a},\mathbf{b})$.

The results of calculations by the expressions (3), (27), (29)-(31) are shown in Figs. 6 and 7.

Figure 6 shows a comparison of the dependences of probabilities $P_{++}(\mathbf{a},\mathbf{b})$ and $P_{+-}(\mathbf{a},\mathbf{b})$ on the angle between polarizers for $\tau = 20$ calculated by the analytical expressions (27), (29)-(31) and obtained by the semiclassical Monte Carlo simulations and processed statistically based on the photonic representations.

Figure 7 shows that the analytical dependences (3), (27), (29) - (31) at $\tau \gg 1$ are close to the predictions of quantum mechanics (2) and (4) but do not match exactly with them. At the same time at $\tau < 1$, the theoretical dependences (3) (27) (29) - (31) are markedly different from the predictions of quantum mechanics (2) and (4) for entangled photons.

Using expressions (3), (27), (29)-(31), one can calculate the parameter (5), the value of which allows judging about possibility to describe the results of quantum experiments using the local hidden-variable theories.

Figure 8 shows the dependence of the parameter $S$, calculated by the expressions ((3), (27), (29)-(31) and (5) for the set orientations $(\mathbf{a},\mathbf{b}) = (\mathbf{a}',\mathbf{b}) = (\mathbf{a}',\mathbf{b}') = 22.5^\circ$ and $(\mathbf{a},\mathbf{b}') = 67.5^\circ$.

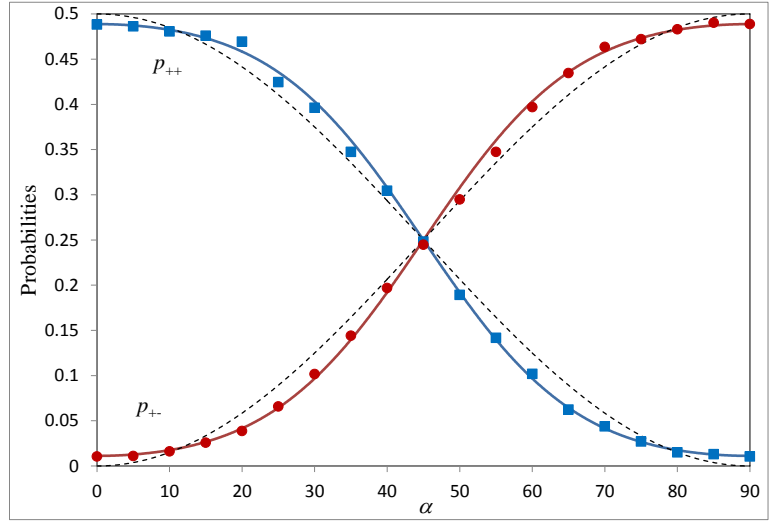

FIG. 6. (Colour online) Comparison of the dependences of probabilities $P_{++}(\mathbf{a},\mathbf{b})$ and $P_{+-}(\mathbf{a},\mathbf{b})$ on the angle between polarizers for $\tau = 20$ calculated by the analytical expressions (27), (29)-(31) (solid lines) and obtained by the semiclassical Monte Carlo simulations (markers); dashed lines are the predictions of quantum mechanics (2).

It also shows the limit values predicted by the local hidden-variable theories $S_{HV} = 2$ and by quantum mechanics $S_{QM} = 2\sqrt{2}$. Calculations show that in the case under consideration, the parameter $S$ has a limit value $S_\infty = 3.2794$ which corresponds to $\tau = \infty$. Thus we see that depending on the nondimensional width $\tau$ of time window, the parameter $S$ can vary from $S \approx 1.4145$ at $\tau \to 0$ up to $S_\infty \approx 3.2794$ at $\tau \to \infty$. In particular, in the semiclassical theory under consideration, the limit value $S_{HV} = 2$ predicted by the local hidden-variable theories is easily exceeded starting from $\tau \approx 0.5$, while at $\tau > 3.7$, the parameter $S$ calculated based on the semiclassical theory exceeds even the limit value $S_{QM} = 2\sqrt{2}$ predicted by quantum mechanics.

Obviously, this fact does not cause much surprise, because the result was obtained within the framework of the classical wave theory of light without using any real particles, and therefore there is no need even to mention a "spooky action at a distance". In this regard, no one will have a desire to construct a hidden-variable theory which would describe the observed coincidences using concept of some fictitious particles (photons).

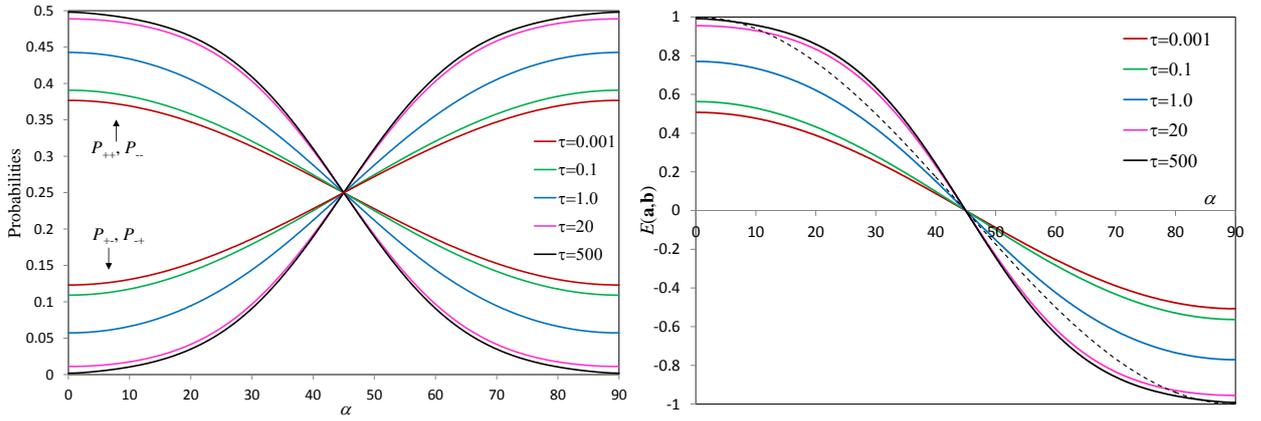

FIG. 7. (Colour online) Theoretical dependences of the probabilities $P_{\pm\pm}(\mathbf{a},\mathbf{b})$ of the pairs of events (left) and correlation coefficient $E(\mathbf{a},\mathbf{b})$ (right) on the angle between polarizers $\alpha$ for different nondimensional width $\tau$ of the time window. The dashed line corresponds to prediction (4) of quantum mechanics.

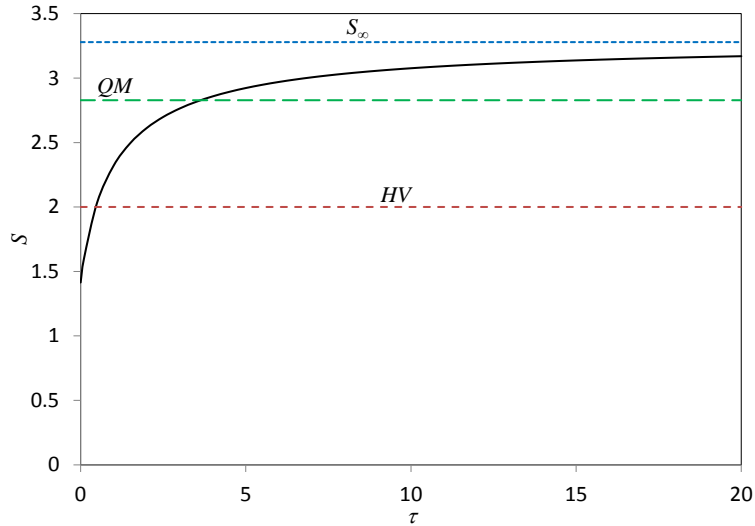

FIG. 8. (Colour online) Theoretical values of the CHSH criterion (5) depending on the nondimensional width $\tau$ of the time window. Dashed lines show the critical values predicted by the local hidden-variable theory (red line, $S_{HV} = 2$) and quantum mechanics for entangled photons (green line, $S_{QM} = 2\sqrt{2}$), as well as the limit value (asymptote) $S_{\infty} = 3.2794$ which corresponds to $\tau = \infty$ (blue line).

## V. HOW CAN ONE OBTAIN EXACTLY THE QUANTUM MECHANICAL PREDICTIONS?

First of all note, that relations

$$P_{\pm\pm}(\alpha = 0) + P_{\pm\pm}(\alpha = \pi/2) = 0.5 \qquad (34)$$

$$P_{++}(\alpha=\pi/2)=P_{--}(\alpha=\pi/2)=P_{+-}(\alpha=0)=P_{-+}(\alpha=0) \quad (35)$$

follow from the theory under consideration, where $P_{\pm\pm}(\alpha=0)$ and $P_{\pm\pm}(\alpha=\pi/2)$ are the probabilities $P_{\pm\pm}(\mathbf{a},\mathbf{b})$ at $\alpha=0$ and $\alpha=\pi/2$, respectively, for a given nondimensional width $\tau$ of the time window. Dependence of probability $P_{++}(\alpha=\pi/2)$ on $\tau$ is shown in Fig. 9.

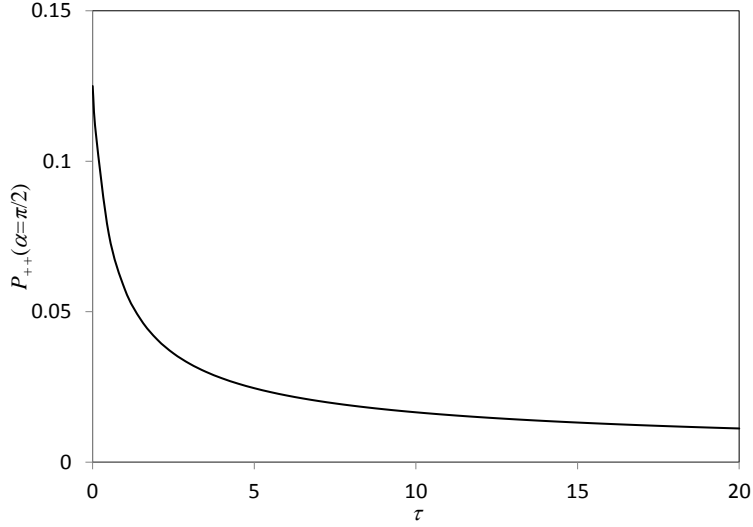

FIG. 9. Dependence of the probability $P_{++}(\alpha=\pi/2)$ on the nondimensional width $\tau$ of the time window.

Taking into account the probabilities (35), one can scale the probabilities $P_{\pm\pm}(\mathbf{a},\mathbf{b})$, using the expressions

$$P'_{++}(\mathbf{a},\mathbf{b})=\frac{P_{++}(\mathbf{a},\mathbf{b})-P_{++}(\alpha=\pi/2)}{1-4P_{++}(\alpha=\pi/2)} \quad (36)$$

$$P'_{--}(\mathbf{a},\mathbf{b})=\frac{P_{--}(\mathbf{a},\mathbf{b})-P_{--}(\alpha=\pi/2)}{1-4P_{--}(\alpha=\pi/2)} \quad (37)$$

$$P'_{+-}(\mathbf{a},\mathbf{b})=\frac{P_{+-}(\mathbf{a},\mathbf{b})-P_{+-}(\alpha=0)}{1-4P_{+-}(\alpha=0)} \quad (38)$$

$$P'_{-+}(\mathbf{a},\mathbf{b})=\frac{P_{-+}(\mathbf{a},\mathbf{b})-P_{-+}(\alpha=0)}{1-4P_{-+}(\alpha=0)} \quad (39)$$

Taking (33) and (35) into account we conclude that the condition (33) is conserved also for scaled probabilities $P'_{\pm\pm}(\mathbf{a},\mathbf{b})$:

$$P'_{++}(\mathbf{a},\mathbf{b})=P'_{--}(\mathbf{a},\mathbf{b}),\ \ P'_{+-}(\mathbf{a},\mathbf{b})=P'_{-+}(\mathbf{a},\mathbf{b}) \quad (40)$$

Taking (32) and (35) into account one obtains

$$P'_{++}(\mathbf{a},\mathbf{b})+P'_{--}(\mathbf{a},\mathbf{b})+P'_{+-}(\mathbf{a},\mathbf{b})+P'_{-+}(\mathbf{a},\mathbf{b})=1 \quad (41)$$

This indicates that the parameters $P'_{\pm\pm}(\mathbf{a},\mathbf{b})$ can also be considered as some probabilities.

As an example, the dependences of the scaled probabilities $P'_{++}(\mathbf{a},\mathbf{b})$ and $P'_{--}(\mathbf{a},\mathbf{b})$ on $\alpha$ are shown in Fig. 10. Ibid, the markers show the predictions of quantum mechanics (2).

We see that at $\tau \geq 1$, the scaled probabilities $P'_{\pm\pm}(\mathbf{a},\mathbf{b})$ differ somewhat from the predictions of quantum mechanics (2), however at $\tau \ll 1$, the scaled probabilities $P'_{\pm\pm}(\mathbf{a},\mathbf{b})$ practically coincides with the quantum mechanical predictions (2). This means that the correlation coefficient (3) and the limit value (7) of the parameter (5) calculated at $\tau \ll 1$ coincide with the predictions of quantum mechanics.

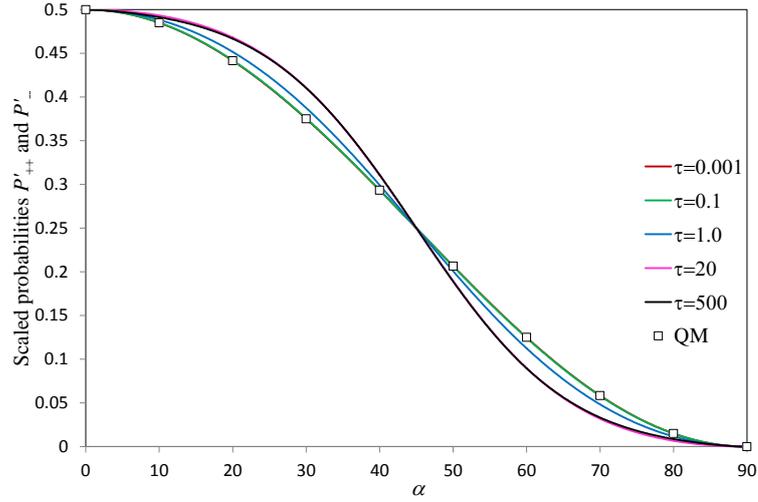

FIG. 10. (Colour online) Dependencies of the scaled probabilities $P'_{++}(\mathbf{a},\mathbf{b})$ and $P'_{--}(\mathbf{a},\mathbf{b})$ on $\alpha$ for different nondimensional width $\tau$ of the time window. Lines $\tau = 0.001$ and $\tau = 0.1$ practically coincide, as well as the lines $\tau = 20$ and $\tau = 500$. Markers correspond to predictions of quantum mechanics (2).

Let us analyze the scaled probabilities $P'_{\pm\pm}(\mathbf{a},\mathbf{b})$.

Taking into account a definition (23), one can write

$$P'_{++}(\mathbf{a},\mathbf{b}) = \frac{N_{++}(\mathbf{a},\mathbf{b}) - N_0 P_{++}(\alpha = \pi/2)}{N'_0} \tag{42}$$

$$P'_{--}(\mathbf{a},\mathbf{b}) = \frac{N_{--}(\mathbf{a},\mathbf{b}) - N_0 P_{--}(\alpha = \pi/2)}{N'_0} \tag{43}$$

$$P'_{+-}(\mathbf{a},\mathbf{b}) = \frac{N_{+-}(\mathbf{a},\mathbf{b}) - N_0 P_{+-}(\alpha = 0)}{N'_0} \tag{44}$$

$$P'_{-+}(\mathbf{a},\mathbf{b}) = \frac{N_{-+}(\mathbf{a},\mathbf{b}) - N_0 P_{-+}(\alpha = 0)}{N'_0} \tag{45}$$

where
$$N'_0 = N_0[1 - 4P_{++}(\alpha = \pi/2)] \tag{46}$$

Let us introduce
$$N'_{++}(\mathbf{a},\mathbf{b}) = N_{++}(\mathbf{a},\mathbf{b}) - N_0 P_{++}(\alpha = \pi/2) \tag{47}$$

$$N'_{--}(\mathbf{a},\mathbf{b}) = N_{--}(\mathbf{a},\mathbf{b}) - N_0 P_{--}(\alpha = \pi/2) \tag{48}$$

$$N'_{+-}(\mathbf{a},\mathbf{b}) = N_{+-}(\mathbf{a},\mathbf{b}) - N_0 P_{+-}(\alpha = 0) \tag{49}$$

$$N'_{-+}(\mathbf{a},\mathbf{b}) = N_{-+}(\mathbf{a},\mathbf{b}) - N_0 P_{-+}(\alpha = 0) \tag{50}$$

Taking (22) and (35) into account, one obtains
$$N'_0 = N'_{++}(\mathbf{a},\mathbf{b}) + N'_{--}(\mathbf{a},\mathbf{b}) + N'_{+-}(\mathbf{a},\mathbf{b}) + N'_{-+}(\mathbf{a},\mathbf{b}) \tag{51}$$

Then for the scaled probabilities $P'_{\pm\pm}(\mathbf{a},\mathbf{b})$, one obtains the definition, similar to definition (23):
$$P'_{\pm\pm}(\mathbf{a},\mathbf{b}) = N'_{\pm\pm}(\mathbf{a},\mathbf{b})/N'_0 \tag{52}$$

Expressions (47)-(52) give us an algorithm for calculation of probabilities $P'_{\pm\pm}(\mathbf{a},\mathbf{b})$:, it is necessary to leave only those time windows at which only one of the detectors behind each polarizer triggered, but at the same time the detectors behind both polarizers triggered simultaneously. As a result, $N_0$ windows which are consistent with the photonic representations will be selected. Further, we assume that there is some background – the random simultaneous triggering the detectors behind different polarizers which are not connected with a "photons hit". This background can be determined by considering the events (in the selected time windows $N_0$) on the detectors $\mathbf{a}_+$ and $\mathbf{b}_+$ at $\alpha = \pi/2$, on the detectors $\mathbf{a}_-$ and $\mathbf{b}_-$ at $\alpha = \pi/2$, on the detectors $\mathbf{a}_+$ and $\mathbf{b}_-$ at $\alpha = 0$ and on the detectors $\mathbf{a}_-$ and $\mathbf{b}_+$ at $\alpha = 0$. Indeed, according to quantum mechanics (2), the probabilities of such events must be equal to zero, and if they are not equal to zero, it should be perceived as the background, which must be rejected. According to (35), all these "background" events have the same probability, therefore, one can consider $P_{++}(\alpha = \pi/2)$ only. Considering that the background does not depend on the angle of the mutual pivot of the polarizers, we should subtract the number of the time windows, in which we expect that the events are connected to the background, from all selected time windows $N_0$ (for given $\tau$). Obviously, the number of such "background" time windows will be equal to $N_0 P_{++}(\alpha = \pi/2)$. As a result, according to (47)-(50), $N'_{\pm\pm}(\mathbf{a},\mathbf{b})$ "good" time windows remain, for which a statistical processing (52) is performed. Thus the conditional probabilities $P'_{\pm\pm}(\mathbf{a},\mathbf{b})$ defined in this manner for $\tau \ll 1$ exactly correspond to predictions of quantum mechanics for so-called "entangled photons"" (Fig. 10).

Let us formalize this analysis for the case $\tau \ll 1$.

Taking (24), (25), (28), (30) and (31) into account, in this case one obtains

$$P_{\pm\pm}(\mathbf{a},\mathbf{b}) = A \langle I_\pm(\mathbf{a}) I_\pm(\mathbf{b}) \rangle \tag{53}$$

where $A$ is the parameter which is defined from the normalization condition (32).

Taking (10) and (12) into account, we write the expression (53) in the form

$$P_{\pm\pm}(\mathbf{a},\mathbf{b}) = A \langle E_\pm^2(\mathbf{a}) E_\pm^2(\mathbf{b}) \rangle \tag{54}$$

where $E_\pm(\mathbf{a})$ are the components of the vector $\mathbf{E}$ of classical electromagnetic (light) wave at the entrance of polarizer $\mathbf{a}$, respectively, in parallel (+) or perpendicular (-) to the axis of polarizer. Let us choose an arbitrary coordinate system $(x, y)$ in which the components of the vector $\mathbf{E}$ we denote as $E_1 = E_x$ and $E_2 = E_y$.

Then for components $E_\pm(\mathbf{a})$ and $E_\pm(\mathbf{b})$ one obtains the expressions

$$E_\pm(\mathbf{a}) = \psi_{\pm i}(\mathbf{a}) E_i \tag{55}$$

$$E_\pm(\mathbf{b}) = \psi_{\pm i}(\mathbf{b}) E_i \tag{56}$$

where $i = 1,2$; summation is carried out by repeated indices, while the parameters $\psi_{\pm i}(\mathbf{a})$ and $\psi_{\pm i}(\mathbf{b})$ are connected with the angles of pivot of the axes of polarizers $\mathbf{a}$ and $\mathbf{b}$ with respect to the axis $x$ of the chosen coordinate system similar to expression (11).

By virtue of isotropy of the system under consideration

$$\langle E_\pm^2(\mathbf{a}) \rangle = \langle E_\pm^2(\mathbf{b}) \rangle = \langle E_x^2 \rangle == \langle E_y^2 \rangle \tag{57}$$

Taking into account (55) and (56), one obtains

$$\langle E_\pm^2(\mathbf{a}) \rangle = \psi_{\pm i}(\mathbf{a}) \psi_{\pm k}(\mathbf{a}) \langle E_i E_k \rangle, \; \langle E_\pm^2(\mathbf{b}) \rangle = \psi_{\pm i}(\mathbf{b}) \psi_{\pm k}(\mathbf{b}) \langle E_i E_k \rangle \tag{58}$$

For the normal distribution (21)

$$\langle E_i E_k \rangle = \delta_{ik} \tag{59}$$

As a result, one obtains

$$\langle E_\pm^2(\mathbf{a}) \rangle = \langle E_\pm^2(\mathbf{b}) \rangle = 1 \tag{60}$$

$$\langle E_\pm^2(\mathbf{a}) \rangle = \psi_{\pm i}(\mathbf{a}) \psi_{\pm i}(\mathbf{a}), \; \langle E_\pm^2(\mathbf{b}) \rangle = \psi_{\pm i}(\mathbf{b}) \psi_{\pm i}(\mathbf{b}) \tag{61}$$

$$\psi_{\pm i}(\mathbf{a}) \psi_{\pm i}(\mathbf{a}) = \psi_{\pm i}(\mathbf{b}) \psi_{\pm i}(\mathbf{b}) = 1 \tag{62}$$

According to (24), the probabilities of single events are determined by the expressions

$$P_\pm(\mathbf{a}) = B \langle E_\pm^2(\mathbf{a}) \rangle; \; P_\pm(\mathbf{b}) = B \langle E_\pm^2(\mathbf{b}) \rangle \tag{63}$$

where the parameter $B$ is determined from the normalization conditions

$$P_+(\mathbf{a}) + P_-(\mathbf{a}) = 1; \; P_+(\mathbf{b}) + P_-(\mathbf{b}) = 1 \tag{64}$$

which follow from the rule of selection of "appropriate" time windows.

Taking (60) and (63) into account, one obtains $B = \frac{1}{2}$, which is equivalent to the result (1) of quantum mechanics.

Taking (61) into account, the expressions (63) can formally be written in form

$$P_\pm(\mathbf{a}) = \frac{1}{2}\psi_{\pm i}(\mathbf{a})\psi_{\pm i}(\mathbf{a}); \ P_\pm(\mathbf{b}) = \frac{1}{2}\psi_{\pm i}(\mathbf{b})\psi_{\pm i}(\mathbf{b}) \quad (65)$$

Let us introduce the functions

$$\Psi_\pm(\mathbf{a}) = \frac{1}{\sqrt{2}}[\psi_{\pm 1}(\mathbf{a}) + i\psi_{\pm 2}(\mathbf{a})], \ \Psi_\pm(\mathbf{b}) = \frac{1}{\sqrt{2}}[\psi_{\pm 1}(\mathbf{b}) + i\psi_{\pm 2}(\mathbf{b})] \quad (66)$$

Then the expressions (65) take the form

$$P_\pm(\mathbf{a}) = |\Psi_\pm(\mathbf{a})|^2; \ P_\pm(\mathbf{b}) = |\Psi_\pm(\mathbf{b})|^2 \quad (67)$$

It follows that the functions $\Psi_\pm(\mathbf{a})$ and $\Psi_\pm(\mathbf{b})$ are the wave functions of single events observed at the detectors $a_\pm$ and $b_\pm$.

Using (55) and (56), the expression (54) takes the form

$$P_{\pm\pm}(\mathbf{a},\mathbf{b}) = A\psi_{\pm i}(\mathbf{a})\psi_{\pm j}(\mathbf{a})\psi_{\pm k}(\mathbf{b})\psi_{\pm m}(\mathbf{b})\langle E_i E_j E_k E_m \rangle \quad (68)$$

For the normal distribution (21) one obtains

$$\langle E_i E_j E_k E_m \rangle = \delta_{ij}\delta_{km} + \delta_{ik}\delta_{jm} + \delta_{im}\delta_{jk} \quad (69)$$

Then

$$P_{\pm\pm}(\mathbf{a},\mathbf{b}) = A[\psi_{\pm i}(\mathbf{a})\psi_{\pm i}(\mathbf{a})\psi_{\pm k}(\mathbf{b})\psi_{\pm k}(\mathbf{b}) + 2\psi_{\pm i}(\mathbf{a})\psi_{\pm i}(\mathbf{b})\psi_{\pm k}(\mathbf{a})\psi_{\pm k}(\mathbf{b})] \quad (70)$$

Taking (62) into account one can write (70) in the form

$$P_{\pm\pm}(\mathbf{a},\mathbf{b}) = A[1 + 2(\psi_{\pm i}(\mathbf{a})\psi_{\pm i}(\mathbf{b}))^2] \quad (71)$$

If the angle between the axes of the polarizers $\mathbf{a}$ and $\mathbf{b}$ is equal to $\alpha = \pi/2$, taking (21) and (54) into account one obtains

$$P_{++}(\alpha = \pi/2) = A\langle E_x^2(\mathbf{a})E_y^2(\mathbf{b})\rangle = A\langle E_x^2(\mathbf{a})\rangle\langle E_y^2(\mathbf{b})\rangle = A \quad (72)$$

Then taking (35)-(39) and (72) into account one obtains

$$P'_{\pm\pm}(\mathbf{a},\mathbf{b}) = A'(\psi_{\pm i}(\mathbf{a})\psi_{\pm i}(\mathbf{b}))^2 \quad (73)$$

where the parameter $A'$ is determined from the normalization condition (41). Taking the properties of the matrix $\psi_{\pm i}$ into account, one obtains $A' = \frac{1}{2}$.

Let us introduce the functions

$$\Psi_{\pm\pm}(\mathbf{a},\mathbf{b}) = \frac{1}{\sqrt{2}}\psi_{\pm i}(\mathbf{a})\psi_{\pm i}(\mathbf{b}) \tag{74}$$

or in expanded form

$$\Psi_{\pm\pm}(\mathbf{a},\mathbf{b}) = \frac{1}{\sqrt{2}}\left[\psi_{\pm 1}(\mathbf{a})\psi_{\pm 1}(\mathbf{b}) + \psi_{\pm 2}(\mathbf{a})\psi_{\pm 2}(\mathbf{b})\right] \tag{75}$$

Taking into account (74), the expression (73) can be written in the form

$$P'_{\pm\pm}(\mathbf{a},\mathbf{b}) = \left|\Psi_{\pm\pm}(\mathbf{a},\mathbf{b})\right|^2 \tag{76}$$

Thus, we have obtained (up to notation) the well-known result of quantum mechanics: the state of the system which is in entangled state, is described by the wave function (75), which cannot be factorized into a product of two states associated to each object, at the same time the probability (76) of realization of any of the possible binary events for such a system is equal to the square of the corresponding wave function (75).

## VI. CONCLUDING REMARKS

Thus we see that the EPRB Gedankenexperiment and entanglement of photons have a simple explanation within the framework of classical electrodynamics and classical optics without involving such a concept as a "photon".

Indeed, there are no particles (photons) in the model under consideration, while light is considered as a classical electromagnetic wave; in this case the discrete events on the detectors (clicks of detectors) are associated not with hitting the particles (photons), but with an excitation of the atoms of the detector by the classical electromagnetic wave according to the relation (8), which is the result of solution of the Schrödinger equation. In this regard, it would be more correct to talk not about the entanglement of photons, but about the entanglement of events for different detectors, or, more precisely, about the correlation of events for different detectors. In particular, if, as was proposed in [21,23], we will call by the photons, not some mythical light particles, but the discrete events of triggering the detectors under the action of classical light wave, we will not face with such paradoxes as the "wave-particle duality" and the "spooky action at a distance".

Thus, as in the case of Hanbury Brown and Twiss effect, the correlation of the events in the EPRB Gedankenexperiment, which is called the entanglement of photons, is connected with the correlation of intensities of classical light waves arriving at the different detectors. The predictions of quantum mechanics for "entangled photons" are adapted to the experimental data at the expense of an artificial rejection of the "bad" events which do not fit into the photonic representations. We note that at processing of the results of real EPRB experiments [5-16], such concepts as the detectors efficiency and the "background events" are actually used; this gives a justification for rejection of the "wrong" time windows and events [24-26]. Therefore, the selection of the "suitable" events for subsequent statistical processing considered in this paper, is fully consistent with the existing practice of processing the results of real EPRB experiments.

Thus we can conclude that a violation of the Bell's inequalities proves only that the intensities of light waves arriving at different detectors are correlated in full compliance with classical electrodynamics and classical optics.

In this connection, a question arises: can the classical "entangled" light waves be used for quantum computing and quantum cryptography? Is it possible to use the classical correlated light beams taking into account their specific character of interaction with the detectors to construct the computational algorithms, similar to "quantum" algorithms?

## ACKNOWLEDGMENTS

Funding was provided by the Tomsk State University competitiveness improvement program.